\documentclass[12pt]{article}
\usepackage{amsmath}
\usepackage{amstext,amssymb,amsfonts, color}
\usepackage{dsfont}
\usepackage{bbm}
\usepackage[dvips]{graphicx}
\usepackage{epsfig}
\usepackage{color}
\usepackage{cite}
\usepackage{epic}

\advance\voffset by -1.5cm
\advance\hoffset by -1.25cm
\definecolor{darkgreen}{rgb}{0,0.6,0}
\def\thefootnote{\fnsymbol{footnote}}

\def\rem#1{}

\newcommand{\N}{{\cal N}}

\newcommand{\nn}{{\nonumber}}

\newcommand{\RR}{{\mathbb R}}

\usepackage{amsmath,amsfonts,epsfig,color,latexsym}
\usepackage{amssymb}
\usepackage[english, USenglish]{babel}
\usepackage{epsfig}
\usepackage{rotating}
\usepackage{slashed}

\numberwithin{equation}{section}

\begin{document}


\thispagestyle{empty}
\renewcommand{\thefootnote}{\fnsymbol{footnote}}

{\hfill \parbox{2.45cm}{
    DESY 09-216 \\ 
}}

\bigskip\bigskip

\begin{center} \noindent \Large \bf
On the Relation between Hybrid and\\ Pure Spinor String Theory
\end{center}

\bigskip\bigskip\bigskip

\centerline{ \normalsize \bf Sebastian Gerigk$^a$ and 
Ingo Kirsch$^b$ \footnote[1]{\noindent \tt email: 
gerigk@phys.ethz.ch, ingo.kirsch@desy.de}} 

\bigskip

\centerline{\it ${}^a$ Institut f\"ur Theoretische Physik, ETH
  Z\"urich}
\centerline{\it 
Wolfgang-Pauli-Strasse 27, CH-8093 Z\"urich, Switzerland}
\bigskip
\centerline{\it ${}^b$ DESY Hamburg, Theory Group,}
\centerline{\it Notkestrasse 85, D-22607 Hamburg, Germany}

\bigskip
\bigskip\bigskip

\renewcommand{\thefootnote}{\arabic{footnote}}

\centerline{\bf \small Abstract}
\medskip

{\small \noindent In this paper we revisit Berkovits' pure spinor
  formalism in lower dimensions. We are particularly interested in
  relating a six-dimensional pure spinor action previously constructed
  in the literature to other superstring formalisms.  In order to gain
  some insight into six-dimensional pure spinors, we first derive
  their action by gauge-fixing the classical six-dimensional
  Green-Schwarz action. We then consider a hybrid pure spinor
  construction in which the spacetime symmetries of six of the ten
  dimensions are described in pure spinor variables, while the
  remaining four dimensions are parameterized in terms of RNS
  variables. We relate this pure spinor formalism to the
  Berkovits-Vafa-Witten hybrid formalism of string theory on $\RR^6
  \times T^4$.}


\setcounter{tocdepth}{2}

\setcounter{equation}{0}
\section{Introduction}

In the past ten years alternative superstring formalisms have been
developed to surmount the shortcomings of the Ramond-Neveu-Schwarz
(RNS) and Green-Schwarz (GS) formalisms.  The most prominent among
them is Berkovits' pure spinor formalism \cite{B1,B2,B3}, see
\cite{Berkovits02, Bedoya:2009np, Oz:2009tb} for reviews and lectures.
As the GS formalism, the pure spinor theory exhibits manifest super
Poincar\'e invariance but in contrast to the former it can be
quantised in a straightforward manner.

Naturally one is also interested in compactifications of the pure
spinor formalism to lower dimensions and their relation to standard
superstring theories. Pure spinor models in $d=2,4,6$ (flat)
dimensions were constructed in \cite{Grassi:2005sb, Wyllard, B3}, see
also \cite{Adam:2006bt} for non-critical pure spinor superstrings. The
$d=10$ pure spinor theory has been related to the RNS superstring by
twisting the ten spin-half RNS fermions using an $SO(10)/U(5)$ pure
spinor variable \cite{Berkovits07}. In the same paper
\cite{Berkovits07} it was also shown that the $d=10$ formalism can be
obtained by gauge-fixing the GS superstring.  Unlike the $d=10$ case,
the relation of lower-dimensional models to known string theories is
rather elusive and has not yet been proven.

In this paper we focus on the $d=6$ pure spinor theory in flat space
\cite{Grassi:2005sb, Wyllard}. In section~\ref{sec2} we show along the
lines of \cite{Berkovits07} that the $d=6$ pure spinor action can be
obtained by gauge-fixing the six-dimensional GS action. The
latter is known to contain $4$ first-class and $4$ second-class
constraints.  After an appropriate splitting of the ghost variables,
the $4$ second-class constraints can be converted into $2$ first-class
constraints, giving $6$ first-class constraints in total.
Gauge-fixing the corresponding Lagrange multipliers will then
introduce six bosonic ghosts. Five of them make up the pure spinor,
which has five independent components in six dimensions. After
removing the sixth ghost by a similarity transformation, the resulting
action becomes the $d=6$ pure spinor action.

In section~\ref{sec3} we then discuss the relation to the
six-dimensional Berkovits-Vafa-Witten hybrid formalism for superstring
theory on $\RR^6 \times T^4$ \cite{BVW}. For this, we supplement the
six-dimensional pure spinor theory of \cite{Grassi:2005sb, Wyllard}
with a four-dimensional action for the compactification on the
four-torus~$T^4$. The internal part on $T^4$ is formulated in RNS
variables and is the same as in the hybrid formalism.  We therefore
only need to show that the external part on $\RR^6$ of the hybrid
action (plus the action for the chiral bosons $\sigma$ and $\rho$) can
be replaced by the six-dimensional pure spinor action of
\cite{Wyllard, Grassi:2005sb}.

\begin{figure}
\begin{center}
\noindent 
\input{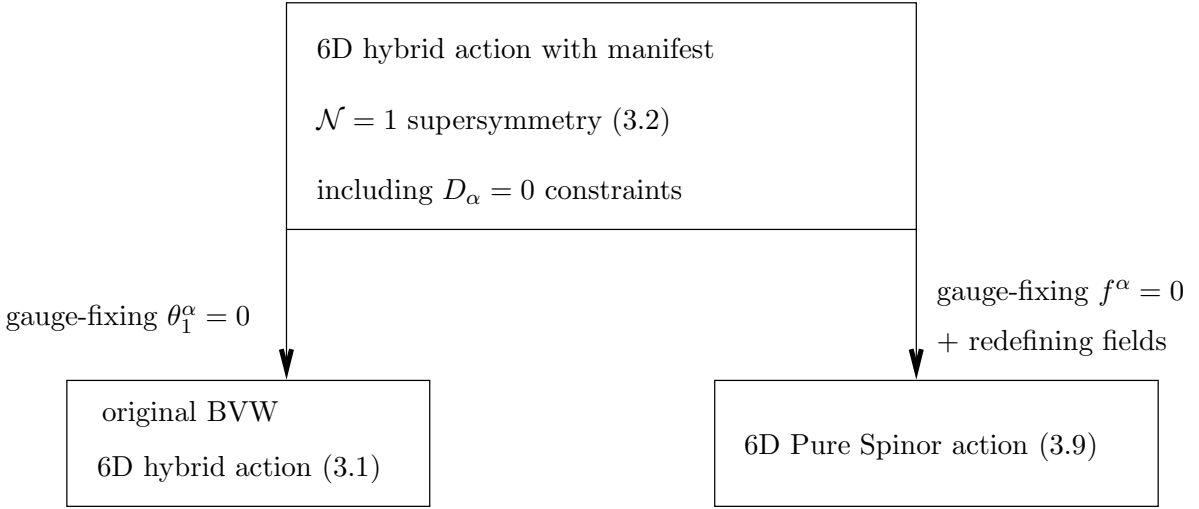}
\end{center}
\caption{\textsl{Schematical illustration of the equivalence of the 
6D hybrid formalism and the 6D pure spinor formalism as introduced in
\cite{Wyllard, Grassi:2005sb}.}} \label{scheme}
\end{figure}

In order to show the equivalence of both formalisms, we perform a
series of manipulations of the original hybrid action.  An overview is
given in figure~\ref{scheme}, which begins with the original hybrid
action (left box) rewritten with manifest six-dimensional
supersymmetry (top box) and ends at the pure spinor action (right
box). In the original hybrid action only half of the $\N=1$ superspace
variables are manifest, {\em i.e.}\ four out of eight $\theta$
variables.  We therefore follow \cite{Berkovits:1999du} and add four
further $\theta$'s and four ``harmonic'' constraints $D_\alpha$ to the
action, giving the hybrid action with manifest $\N=1$ supersymmetry
(top box). These constraints are first-class and allow the additional
$\theta$'s to be gauged away \cite{Berkovits:1999du}. Instead of
gauging them away, we add the four constraints $D_\alpha$ to the
action using Lagrange multipliers~$f^\alpha$. We then gauge-fix the
$f^\alpha$ to zero, which introduces four bosonic ghost fields.
Another ghost field of the same kind comes from the
$\sigma$-$\rho$-sector of the hybrid action.  After an appropriate
field redefinition, the four plus one ghost fields combine to give the
five components of the six-dimensional pure spinor, and the hybrid
action turns into the pure spinor action (right box). We also show
that the same field transformation maps the BRST operators into each
other. We close with some remarks on the cohomology.

\setcounter{equation}{0}
\section{Green-Schwarz versus pure spinor formalism in six dimensions}
\label{sec2}

In this section we derive the pure spinor action in six dimensions by
gauge-fixing the six-dimensional Green-Schwarz action, which is
consistent on the classical level. The impatient reader, who is
interested only in the connection to the hybrid string, may wish to
immediately proceed to section~\ref{sec3} after reading the general
introduction on pure spinors in six dimensions in section~\ref{sec21}.

\subsection{Pure spinors in six-dimensions}\label{sec21}

Pure spinors in six dimensions are discussed in detail in
\cite{Grassi:2005sb, Wyllard}. Here we only review some of their basic
properties. Six-dimensional pure spinors are given by two $SO(6)$ Weyl
spinors $\lambda_I^\alpha$ ($I=1,2$; $\alpha=1,...,4$) which are
subject to the constraint
\begin{align}
\varepsilon^{IJ} \lambda_I^\alpha \Gamma^m_{\alpha\beta} \lambda_J^\beta = 0
\,, \label{PSconstraint}
\end{align}
where the $\Gamma^m$ are the antisymmetric $4\times 4$ off-diagonal
blocks of the six-dimensional gamma matrices given in
appendix~\ref{appA}. They satisfy 
\begin{equation}
\Gamma^{(m}_{\alpha\beta} \Gamma^{n)\beta\gamma}= \eta^{mn}
\delta_\alpha^\gamma\,. \label{Clifford}
\end{equation}
The corresponding matrices with upper indices are defined by
$\Gamma^{m\alpha\beta}=\frac{1}{2}\varepsilon^{\alpha\beta\gamma\delta}
\Gamma^m_{\gamma\delta }$.

We may now go to light-cone gauge by defining $\Gamma^\pm=\Gamma^0 \pm
\Gamma^5$. This induces a symmetry breaking of $SO(6)$ down to $SO(4)$
under which the Weyl spinor representation $\bf{4_s}$ decomposes into
${\bf 2_s} \oplus {\bf 2_c}$. Explicitly, the spinor can be solved by
writing $\lambda_I^\alpha = (\lambda_I^A,\lambda_I^{\dot A})$ with $A,
\dot A=1,2$. The pure spinor constraint (\ref{PSconstraint}) then
decomposes into
\begin{align}\label{constraints}
  \varepsilon^{IJ} \lambda_I^A \varepsilon_{AB} \lambda_J^B =
  0\,,\qquad 
  \varepsilon^{IJ} \lambda_I^{\dot A} \varepsilon_{\dot A
    \dot B} \lambda_J^{\dot B} = 0 \,,\qquad
  \varepsilon^{IJ} \lambda_I^{A} \lambda_J^{\dot B} = 0
\,.
\end{align}
As shown in \cite{Grassi:2005sb}, the second and third constraints are
solved by
\begin{align}\label{gs1}
\lambda^{\dot B}_I = h^{\dot B}_{(0)}{}_A \lambda^A_I \,
\end{align}
provided the first constraint is satisfied. The field $h^{\dot
B}_{(0)}{}_A$ consists of four degrees of freedom. However, there is
an infinite number of gauge symmetries ($n=0,1,...$)
\begin{align}\label{gs2}
\delta h^{\dot B}_{(n)}{}_A = \varepsilon^{IJ} \varepsilon_{AB}
\eta^{\dot B}_{(n)I} \lambda^B_J \,,\qquad
\delta \eta^{\dot B}_{(n)I} = h^{\dot B}_{(n+1)}{}_A \lambda^A_I \,,
\end{align}
which reduce the number of degrees of freedom of $\lambda^{\dot B}_I $
to $4-4+4-4+...= 4 \sum_k (-1)^k= 2$.  At the same time the first
constraint reduces the number of degrees of freedom of $\lambda^A_I$
to three. The six-dimensional pure spinor therefore has only five
independent components.

\subsection{Equivalence of the pure spinor formalism and the Green-Schwarz
action in six dimensions}

For $\N = (1, 0)$ supersymmetry in $d=6$ the left-moving (holomorphic)
matter worldsheet fields are $(x^m, \theta_I^\alpha, p_\alpha^I)$,
where $\theta_I^\alpha$ is a doublet $(I = 1, 2)$ of four-component
Weyl spinors, and $p^I_\alpha$ are their conjugate momenta
$(\alpha=1,...,4)$.

The six-dimensional Green-Schwarz action in first-order form is given
by
\begin{align}
S=\int d^2z \left[ \frac{1}{2} \partial x^m \bar \partial x_m +
p^I_\alpha \bar \partial \theta_I^\alpha - e (\frac{1}{2} \partial x^m 
\partial x_m+ p^I_\alpha \partial \theta_I^\alpha) \right] 
\label{GSaction} \,,
\end{align}
where $m=0,...,5$. Since the action is in conformal gauge, the
Virasoro constraint $T=0$ has been added to the action using a
Lagrange multiplier $e$. The action is also supplemented by the
constraints
\begin{align}
d_\alpha^I = p^I_\alpha -\frac{1}{2} \varepsilon^{IJ} (\Gamma_m \theta_J)_\alpha
\big(\partial x^m + \frac{1}{4} 
\varepsilon^{KL} (\theta_K \Gamma^m \partial \theta_L)\big)
\label{constraintGS} \, .
\end{align}
Setting $d_\alpha^I=0$ and substituting the resulting equation for the
conjugate momentum $p^I_\alpha$ back into (\ref{GSaction}) yields the
standard form of the Green-Schwarz action.

Using the identity (\ref{identity}), the constraints
(\ref{constraintGS}) can be rewritten as
\begin{equation} \label{GScon}
d_\alpha^I = p^I_\alpha -\frac{1}{2} \varepsilon^{IJ} (\Gamma_m \theta_J)_\alpha
\partial x^m - \frac{1}{4} \varepsilon^{IJ} 
\varepsilon^{KL} \varepsilon_{\alpha\beta\gamma\delta}
\theta^\beta_J\theta^\gamma_K \partial \theta^\delta_L \,.
\end{equation}
They satisfy the OPE \cite{Grassi:2005sb}
\begin{equation}
d_{I\alpha}(z)d_{J\beta}(w) = -(z-w)^{-1} \varepsilon_{IJ}
\slashed{\Pi}_{\alpha\beta} \,,
\end{equation}
where $\slashed{\Pi} = \Pi_m \Gamma^m$. Since $\slashed{\Pi}^2=0$,
$d_\alpha^I$ separate into four first-class and four second-class
constraints, denoted by $d^I_A$ and $d^I_{\dot A}$, respectively.

We now incorporate the GS-constraints (\ref{GScon}) into the action
(\ref{GSaction}) by introducing Lagrange multiplier terms $f^\alpha_I
d^I_\alpha$. The first-class constraints can be eliminated by
gauge-fixing $f_I^A = 0$ in $f^\alpha_I d^I_\alpha=f^A_I d^I_A
+f^{\dot A}_I d^I_{\dot A}$.  This introduces four
$\beta$-$\gamma$-systems with weights $(1, 0)$, denoted by $\beta^I_A$
and $\gamma^A_I$. Introducing also the usual gauge-fixing term $b\bar
\partial c$ for $e=0$, we get
\begin{align} \label{actionint}
 S= \int d^2z \left[ \frac{1}{2} \partial x^m \bar \partial x_m +
    p^I_\alpha \bar \partial \theta_I^\alpha + f^{\dot A}_I d^I_{\dot
      A} + b \bar \partial c + \beta^I_A \bar \partial \gamma^A_I \right] \,.
\end{align}

Next, we need to express the four second-class constraints $d^I_{\dot
A}$ in terms of two first-class constraints. For this, we regroup the
ghosts $\gamma^A_I \rightarrow (\tilde \gamma, \lambda^A_I)$ into
$\lambda^A_I$, which is subject to the first constraint of
(\ref{constraints}) and therefore has three independent components,
and one component~$\tilde \gamma$. We then define the first-class
constraints
\begin{align} \label{HM}
H^M \equiv \varepsilon^{IJ} \lambda^\alpha_I 
(\Gamma^- \Gamma^M)_{\alpha\dot \alpha} 
d_J^{\dot \alpha} =
\varepsilon^{IJ} \lambda^A_I \sigma^M_{A\dot A} d_J^{\dot A}
\end{align}
with $M=1,2,3,4$. Here we used that the matrices $\Gamma^-
\Gamma^M$ are $4 \times 4$ are matrices of the type
\begin{align}
\begin{pmatrix}
0 & \sigma^M\\ 0 & 0
\end{pmatrix}\,,
\end{align}
where $\sigma^M$ are the standard $2\times 2$ Pauli matrices (with one
of them the identity operator~$\mathbf{1}_2$). As the four
$\sigma^M$-matrices give a basis of $2\times 2$ matrices, the
constraints in Eq.~(\ref{HM}) are equivalent to the four constraints
\begin{equation}
 \epsilon^{IJ}\lambda^A_I d^{\dot{A}}_J = 0 \,.
\end{equation}
These can be solved by
\begin{align}
d_J^{\dot A}= M_{(0)\dot B}^{\dot A}
\lambda_J^{\dot B}
\end{align} for any linear mapping $M$, which has four degrees of freedom. 
However, as in (\ref{gs1}) and (\ref{gs2}), there is an infinite
number of gauge symmetries which reduce the number of degrees of
freedom to two. Thus, only two of the four constraints $H^M$
are independent.

We may therefore write $f^{\dot A}_I d^I_{\dot A}=h_M H^M$, where only
two of the four Lagrange multipliers $h_M$ are non-vanishing, say
those for $M=0,1$. Gauge-fixing them to zero yields two further
bosonic $\beta$-$\gamma$-systems which we denote by $w^M$ and
$\lambda_M$ (now e.g.\ $M=0,1$ only). The action is then
\begin{align}
  S= \int d^2z \left[ \frac{1}{2} \partial x^m \bar \partial x_m +
    p^I_\alpha \bar \partial \theta_I^\alpha + w^M \bar \partial 
  \lambda_M + b \bar
    \partial c + w^I_A \bar \partial \lambda^A_I + \tilde \beta 
    \bar \partial \tilde \gamma  \right] \,.
\end{align} 
Here the last two terms descend from the last term in
(\ref{actionint}) according to the decompositions $\gamma^A_I
\rightarrow (\tilde \gamma, \lambda^A_I)$ and $\beta_A^I \rightarrow
(\tilde \beta, w_A^I)$. 

The spinors $\lambda^M$ (two degrees of freedom) and $\lambda^A_I$
(three degrees of freedom) then make up the pure spinor
$\lambda^\alpha_I$ (five independent components). Note that the two
degrees of freedom of $\lambda^M$ can be rearranged into the pure
spinor components $\lambda_I^{\dot A}$, which has also two independent
components, as shown in section~\ref{sec21}. As in the ten-dimensional
case \cite{Berkovits07}, an appropriate similarity transformation of
the action removes the ghost terms $b \bar \partial c$ and $\tilde
\beta \bar \partial \tilde \gamma$. The action can then be written in
terms of the pure spinor $\lambda^\alpha_I$ as
\begin{align}
  S= \int d^2z \left[ \frac{1}{2} \partial x^m \bar \partial x_m +
    p^I_\alpha \bar \partial \theta_I^\alpha + w^I_\alpha \bar \partial 
    \lambda^\alpha_I \right] \,, \label{PSaction}
\end{align}
which is the six-dimensional pure spinor action of
\cite{Grassi:2005sb, Wyllard}.

Let us finally determine the central charges. The pure spinor fields
$w, \lambda$ formally represent five (bosonic)
$\beta$-$\gamma$-systems with weights $(1, 0)$. The fermions $p,
\theta$ correspond to eight (fermionic) $b$-$c$-systems of weight
$(1,0)$. The central charges are therefore $c_{w,\lambda}=5\cdot
2=10$, $c_{p,\theta}=8\cdot (-2)=-16$ and $c_x=6$.  The total central
charge is thus zero.

\setcounter{equation}{0}
\section{From hybrid to pure spinor formalism} \label{sec3}

In this section we derive the pure spinor action in six dimensions
\cite{Wyllard, Grassi:2005sb} from the Berkovits-Vafa-Witten hybrid 
formalism for string theory on $\RR^6 \times T^4$ \cite{BVW}. More
precisely, we show that the external part on $\RR^6$ of the hybrid
action (plus the action for the bosons $\rho$ and $\sigma$) can
be replaced by the six-dimensional pure spinor action found in
\cite{Wyllard, Grassi:2005sb}. The internal part on $T^4$ is described
in RNS variables and remains the same in the pure spinor formalism.

We proceed as outlined in the introduction and summarized in
figure~\ref{scheme}. In section~\ref{hfsusy} we rewrite the $\N=2$
hybrid action such that ${\cal N} = 1$ supersymmetry becomes
manifest~\cite{Berkovits:1999du}. In section~\ref{hfgaugefixed} we
discuss the corresponding gauge-fixed action. In section~\ref{equival}
we will then show that, after an appropriate field redefinition, the
gauge-fixed hybrid action turns into the six-dimensional pure spinor
action. Finally, in section~\ref{BRSTop} we relate the corresponding
BRST operators.

\subsection{Hybrid formalism with manifest $\N=1$ superspace variables}
\label{hfsusy}

The (holomorphic part of the) hybrid action in its original form is
given by \cite{BVW}
\begin{align} \label{hybrid1}
S_{\rm hybrid}= \int d^2 z\left[\frac{1}{2} \partial x^m \bar \partial x_m+
p_\alpha \bar \partial \theta^\alpha \right] + S_B + S_C\,,
\end{align}
where the external part on $\RR^6$ is described by the six bosons
$x^m$ ($m=0,...,5$), four fermions $\theta_\alpha$ and their conjugates
$p^\alpha$ ($\alpha=1,...,4$) and an action $S_B$ for the two chiral bosons
$\sigma$ and $\rho$ appearing in the hybrid formalism. The action
$S_C$ describes the internal part on $T^4$ in RNS
variables. Here we have assumed that the reader is
familiar with the hybrid formalism \cite{BVW}.  A summary of the
fields and their properties is given in appendix~\ref{appB}.

Unlike in the Green-Schwarz formalism, only half of the usual eight
$\theta^\alpha_I$ variables, say $\theta^\alpha=\theta^\alpha_2$, of
six-dimensional $\N=1$ supersymmetry are manifest.  However, as
suggested in \cite{Berkovits:1999du, Berkovits:1999xv}, it is possible
to add $\theta^\alpha_1$ to the hybrid variables, as well as
constraints $D_\alpha$, which allow the additional variables
$\theta^\alpha_1$ to be gauged away.

Then, the hybrid action may be written as
\begin{align} \label{hybrid2}
S_{\rm hybrid}= \int d^2 z \,\left[\frac{1}{2} \partial x^m \bar
\partial x_m + p^I_{\alpha} \bar \partial \theta^{\alpha}_I
\right] + S_B + S_C \,.
\end{align}
Equivalence to the original hybrid action (\ref{hybrid1}) requires that
$\theta^\alpha_1$ and $p_{\alpha}^1$ satisfy the first-class
constraints \cite{Berkovits:1999xv}
\begin{align}
D_\alpha=d^1_\alpha - e^{-\rho - i \sigma} d_\alpha^2 = 0 \,, 
\label{constraint1}
\end{align}
where $x^{\alpha\beta}=(\Gamma_m)^{\alpha\beta} x^m$ and $d_\alpha^I$ defined as before. Since
$D_\alpha(z) \theta^{\beta}_1(w)\sim\delta^\beta_\alpha(z-w)^{-1}$,
the additional variables $\theta^\alpha_1$ transform as
\begin{align}
\delta \theta^\alpha_1(w) = \oint dz\,\varepsilon^\beta(z) D_\beta(z) 
\theta^\alpha_1(w) = \varepsilon^\alpha(w)
\end{align}
under the gauge invariance generated by $D_\alpha$, as required for
superspace variables. The gauge invariance may be used to gauge-fix
$\theta^\alpha_1 = 0$, in which case (\ref{hybrid2}) reduces to
(\ref{hybrid1}). Note that gauge fixing $\theta^\alpha_1 = 0$ does not
produce any ghosts since the generating algebra has trivial
anticommutation relations.

\subsection{Gauge-fixed hybrid action}
\label{hfgaugefixed}

We now implement the constraints of (\ref{constraint1}) into the action
(\ref{hybrid2}) by introducing Lagrange multipliers $f^\alpha$. The
extended action then has the form
\begin{equation} \label{extendedaction}
 S_\text{hybrid}= \int d^2 z\left[\frac{1}{2} \partial x^m \bar \partial x_m 
 +p^I_{\alpha} \bar \partial \theta^{\alpha}_I 
 - f^\alpha D_\alpha \right] + S_B + S_C \,
\end{equation}
with $D_\alpha$ as in (\ref{constraint1}). Both the constraints
$D_\alpha$ and Lagrange multipliers $f^\alpha$ are fields of conformal
weight~1.

On general grounds it can be shown that every constraint induces a
gauge symmetry on the extended action \cite{Henneaux:1992ig}. This
gauge symmetry is given by
\begin{equation}
 \delta_D F(w) = \oint_{C_w} dz\, \epsilon^\alpha(z) D_\alpha (z) F(w)
\end{equation}
for any field $F$ in (\ref{extendedaction}) except for the Lagrange
multipliers $f^\alpha$. The gauge transformation acting on $f^\alpha$
can then be defined such that $S_\text{hybrid}$ is gauge
invariant. The general form of $\delta_D f^\alpha$ can be found in
\cite{Henneaux:1992ig}. Since the constraints anticommute with each
other, $\delta_D f^\alpha$ simplifies a lot to
\begin{equation}
 \delta_D f^\alpha = \bar\partial \epsilon^\alpha \,.
\end{equation}

We now gauge-fix this symmetry such that $f^\alpha=0$.\footnote{Note
that this gauge-fixing is different from that of the previous
subsection, $\theta^1=0$, which led back to (\ref{hybrid1}).} By the
usual Faddeev-Popov method the resulting functional determinant
$\Delta_{FP}=\det(\bar\partial\delta^2(z-w)\delta^\alpha_\beta)$ can
be rewritten as a functional integral over ghost fields, here
$\beta_\alpha$ and $\gamma^\alpha$.  The full action then reads
\begin{equation}
S_{\rm hybrid}=\int d^2z \left[\frac{1}{2} \partial x^m \bar \partial x_m +
p_{\alpha}^I \bar \partial \theta^{\alpha}_I + \beta_\alpha \bar
\partial \gamma^\alpha \right]
+ S_B+ S_C
\,.\label{fullaction}
\end{equation}
The fields $\beta_\alpha$ and $\gamma^\alpha$ are bosonic ghosts of
conformal weight 1 and 0, respectively, and transform in the Weyl
representation ${\bf 4_s}$ of $SO(6)$.

\subsection{Equivalence of the pure spinor and hybrid actions}
\label{equival}

The next step will be to relate the hybrid action in its gauge-fixed
form (\ref{fullaction}) to the pure spinor action
\begin{align}
  S_{\rm ps}&= \int d^2z \left[ \frac{1}{2} \partial x^m \bar \partial x_m +
    p^I_\alpha \bar \partial \theta_I^\alpha + w^I_\alpha \bar \partial 
    \lambda^\alpha_I \right] +S_C \,\label{PSaction2} \,,
\end{align} 
which corresponds to the six-dimensional pure spinor action
(\ref{PSaction}) plus a compact part, $S_C$, in RNS variables.

The pure spinor constraints in six dimensions require two independent
spinors, each in the ${\bf 4_s}$ of $SO(6)$ \cite{Grassi:2005sb,
Wyllard}. For the following it is convenient to temporarily break
$SO(6)$ down to $U(3)$ such that 
\begin{equation}
{\bf 4_s} \oplus {\bf 4_s} \rightarrow {\bf 3} \oplus {\bf 3} \oplus {\bf 1}
\oplus {\bf 1} \,.
\end{equation}
We can write this decomposition under the subgroup $U(3)$ explicitly
as $\lambda^\alpha = (\lambda^+, \lambda^a)$ with $a=1,2,3$. As it is
shown in \cite{Wyllard}, the pure spinor constraint
(\ref{PSconstraint}) implies
\begin{equation}
\lambda^a_2 = \frac{\lambda^+_2}{\lambda^+_1}\lambda^a_1 
\label{constraint3}\,.
\end{equation}
Therefore one of the ${\bf 3}$ representations is completely
determined by the remaining ${\bf 3} \oplus {\bf 1} \oplus {\bf 1}$
representation, which in turn can be interpreted as the $U(3)$
invariant representation of pure spinors. This gives exactly five
degrees of freedom for the pure spinor, as required.

Using the explicit solution (\ref{constraint3}) of the pure spinor
constraint in terms of $U(3)$-invariant representations, the pure
spinor part of the action (\ref{PSaction2}) can solely be expressed in
terms of the five pure spinor degrees of freedom $\lambda_1^+,
\lambda_1^a$ and $\lambda_2^+$ as
\begin{align}\label{314}
 \omega^I_\alpha \bar \partial \lambda^\alpha_I
 = \left(\omega^1_\alpha +
\frac{\lambda^+_2}{\lambda^+_1}\omega^2_\alpha\right)\bar \partial
\lambda^\alpha_1+\left(\lambda_1^\alpha \omega^2_\alpha\right)\bar \partial
\left(\frac{\lambda^+_2}{\lambda^+_1}\right)\,.
\end{align}

\medskip
We may now deduce the hybrid action in the form (\ref{fullaction})
from the pure spinor action (\ref{PSaction2}) using (\ref{314}). It is
natural to assume that four of the five pure spinor degrees of
freedom, $\lambda_1^\alpha$, are related to the four ghosts
$\gamma^\alpha$ in the hybrid formalism. We therefore set
\begin{eqnarray}
\gamma^\alpha = \lambda_1^\alpha\,,\qquad
\beta_\alpha &=& \omega^1_\alpha +
\frac{\lambda^+_2}{\lambda^+_1}\omega^2_\alpha 
\end{eqnarray}
and the pure spinor action simplifies to
\begin{align}
  S_{\rm ps}&= \int d^2z \left[ \frac{1}{2} \partial x^m \bar \partial x_m +
    p^I_\alpha \bar \partial \theta_I^\alpha + \beta_\alpha \bar \partial 
    \gamma^\alpha + w \bar \partial \lambda \right] +S_C \,.
\label{PSaction3}
\end{align}
The variables $\lambda$ and $w$ correspond to the fifth component of
the pure spinor and its conjugate momentum and are defined by $\lambda
\equiv \frac{\lambda^+_2}{\lambda^+_1}$ and $w \equiv \lambda_1^\alpha
\omega^2_\alpha$.

While the first three terms in (\ref{PSaction3}) already agree with
those of (\ref{fullaction}), the fifth component of the pure spinor
still needs to be related to the hybrid variables. A single component
of the pure spinor $w\lambda$-system has central charge $c=2$. In the
hybrid formalism there are two bosonic ghosts $\sigma$ and $\rho$ with
total central charge $c_\sigma+c_\rho= -26+ 28 =2$. We may therefore
conjecture that $\sigma$ and $\rho$ make up the fifth component of the
pure spinor. Indeed, as we will show now, they can be obtained by
bosonising the $w\lambda$-system in the appropriate way.

The $w\lambda$-system is formally a bosonic ghost system with
conformal weights $[\lambda]=0$ and $[w]=1$ and energy-momentum tensor
\begin{align}\label{enps}
T^{w\lambda}=(\partial w) \lambda - \partial (w \lambda) \,.
\end{align}
Each bosonic ghost system can be decomposed into a
free boson, which in the following we denote by $\rho$, and an
anticommuting $bc$ system.  In particular, we may rewrite $w$ and $\lambda$
as
\begin{align}
w=e^{\rho} \partial c \,,\qquad \lambda=e^{-\rho} b \,.
\end{align}
In order to identify $\rho$ with the corresponding ghost in the hybrid
formalism, we choose $Q_\rho=3$ as the background charge for $\rho$
such that its central charge becomes $c_\rho=1+3Q_\rho^2=28$, as
required.  Since, in general, $[e^{n\rho}]=-\frac{n^2}{2} + 
\frac{n}{2}Q_\rho$ for the conformal weight of $e^{n\rho}$, we get
$[e^{\rho}]=1$ and $[e^{-\rho}]=-2$, and therefore $[b]=2$ and $[c]=-1$,
which are the conformal weights of a standard $bc$ system.

As usual, one can go further and bosonise the $bc$ system as
$b=e^{-i\sigma}$ and $c=e^{i\sigma}$, with central charge given by
$c_\sigma=1-3Q_\sigma^2=-26$ for $Q_\sigma=3$. The conformal weights
are $[e^{i n\sigma}]=\frac{n^2}{2} - \frac{n}{2}Q_\sigma$. The
energy-momentum tensor (\ref{enps}) can then be rewritten in terms of
$\rho$ and $\sigma$ as
\begin{align}
T^{\rho,\sigma}= -\frac{1}{2} \partial \rho \partial \rho 
-\frac{1}{2} \partial \sigma \partial \sigma - \frac{3}{2} \partial^2 
(\rho + i \sigma) \,, 
\end{align}
which is identical to the energy-momentum tensor corresponding to the
ghost action $S_B$ in the hybrid formalism \cite{BVW}. The term $\int
d^2z \,w\bar \partial \lambda$ in (\ref{PSaction3}) is therefore
identical to the ghost action $S_B$ in the hybrid formalism. The
action (\ref{PSaction3}) is thus equivalent to the hybrid action
(\ref{fullaction}), $S_{\rm ps}=S_{\rm hybrid}$.

\subsection{Comments on the BRST operators} \label{BRSTop}

As a critical $\N=2$ theory, the hybrid string can be reformulated as
a $\N=4$ topological string theory \cite{BVW, Theisen}. Recall that
every $\N=2$ superconformal theory with central charge $c=6$ gives
rise to a critical $\N=4$ superconformal field theory.  The
corresponding $\N=4$ algebra is generated by the energy momentum
tensor $T$, four fermionic currents $G^\pm$ and $\widetilde G^\pm$,
and three $SU(2)$ currents $J^a$ ($a=1,2,3$). These currents can be
defined from the $\N=2$ currents $[T, G^+, G^-, J]$ by $T$, $G^+, G^-,
\widetilde G^+ \equiv [e^{-\int J}, G^+],\widetilde G^- \equiv
[e^{\int J}, G^-]$, and $J$, $e^{\int J}$, $e^{-\int J}$.  Explicit
expressions for these generators with manifest six-dimensional
superspace variables can be found in \cite{Berkovits:1999du}.

Open $\N=4$ string physical vertex operators in hybrid string theory
satisfy the physical state conditions\footnote{$G_0^+$ is the charge
(zero mode) corresponding to the current $G^+$, etc.}${}^{,}$\footnote{Analogous
conditions hold for the closed
superstring~\cite{BVW}.}
\begin{align}\label{psc} G_0^+ \Phi =
\widetilde G_0^+ \Phi = (J_0-1)\Phi = 0
\,,\qquad \delta \Phi = G_0^+
\widetilde G_0^+ \Lambda^- \,.
\end{align}
Since the cohomology of $\widetilde G_0^+$ is trivial \cite{BVW,
  Theisen}, $\Phi$ can be written as
\begin{align}
\Phi= \widetilde G^+_0 V \,,\qquad G_0^+ \widetilde G_0^+ V = J_0 V = 0
\end{align}
with gauge invariance $\delta V = G_0^+ \Lambda + \widetilde G_0^+ 
\widetilde \Lambda$. This gauge invariance can be fixed such that
\begin{align}
G^-_0 \Phi = \widetilde G^-_0 \Phi = T_0 \Phi  = 0
\end{align}
is automatically satisfied \cite{BVW}. 

We now need to take into account that we introduced additional
$\theta$ variables and added the Lagrange multiplier term $f^\alpha
D_\alpha$. The gauge-fixing $f^\alpha=0$ requires us to impose a
further condition on the physical states. Since $D_\alpha(z)
D_\beta(w)\sim 0$, the gauge symmetries generated by $D_\alpha$ are
abelian such that the additional condition has the simple form
$\gamma^\alpha D_\alpha$.  After gauge-fixing, a physical state must
also satisfy
\begin{align} \label{Qhyb}
Q_\text{hybrid} \Phi = 0 \,,
\end{align}
where
\begin{align}
 Q_\text{hybrid} &= \oint dz \left( \gamma^\alpha D_\alpha \right)
 \,\nonumber\\
&= \oint dz \, \left( \gamma^\alpha d_\alpha^1 - e^{-\rho - i \sigma}
\gamma^\alpha d^2_\alpha  \right) \,. \label{320}
\end{align}
In the second line we used the definition of $D_\alpha$ given by
(\ref{constraint1}).  For the following it is useful to change the
minus sign in (\ref{320}) into a plus sign by exploiting the symmetry
$\theta_2 \rightarrow -\theta_2$ and $d^2 \rightarrow -d^2$, see
\cite{Wyllard}.

\medskip The operator (\ref{320}) can be shown to be equivalent to the
BRST operator of the pure spinor formalism. For that, we use the field
redefinition of the previous subsection,
\begin{align}
\lambda_1^\alpha = \gamma^\alpha\,,\qquad \lambda=\lambda_2^+/\lambda_1^+
=e^{-\rho-i\sigma} \,,
\end{align}
or, equivalently, by Eq.~(\ref{constraint3}),
\begin{eqnarray}
 \lambda^\alpha_1 &=& \gamma^\alpha \nonumber \,,\\
 \lambda^\alpha_2 &=& e^{-\rho - i \sigma} \gamma^\alpha \,.
\label{identification}
\end{eqnarray}
Substituting this into (\ref{320}), 
we get
\begin{equation}\label{325}
 Q_{\rm hybrid} = \oint dz \, \lambda^\alpha_I d^I_\alpha = Q_{\rm ps}
 \, ,
\end{equation}
which is exactly the pure spinor BRST operator $Q_{\rm ps}$, as
defined in \cite{Grassi:2005sb, Wyllard}.  Recall that nilpotence of
$Q_{\rm ps}$ is ensured by the pure spinor constraint
(\ref{PSconstraint}) \cite{Grassi:2005sb, Wyllard}.

\medskip We close with a few comments on the vertex operators in
both theories. Let us restrict to the massless open string vertex
operator which is independent of the compactification variables. In
hybrid string theory this operator is obtained by solving the physical
state conditions (\ref{psc}) {\em and} the `harmonic' BRST-like
condition (\ref{Qhyb}). As found in \cite{Berkovits:1999du,
Berkovits:1999xv}, such an operator describes the six-dimensional {\em
on-shell} degrees of freedom of six-dimensional $\N=1$ super Yang-Mills
theory. The corresponding integrated vertex operator is given
by\footnote{The last term involving the ghost $\gamma^\alpha$ and its
conjugate $\beta_\alpha$ was later added in \cite{B1}, see footnote~3
therein.}
\begin{align}
\Phi_{\rm hybrid}  = \int dz \left[ \Pi^m A_m + \partial \theta^\alpha_I
A_\alpha^I+ d_\alpha^I W^\alpha_I 
+ \frac{1}{2} (\gamma^\alpha (\Gamma^{mn})_\alpha{}^\beta \beta_\beta) F_{mn} 
 \right]
\end{align}
where $\Pi^m$ are the superspace momenta, $W^\alpha_I$ the superspace
spinor field strengths and $A_m$, $A_\alpha^I$ the superspace gauge
fields \cite{Berkovits:1999du, Berkovits:1999xv}. $F_{mn}$ is a
superspace field strength whose lowest component is the gluon field
strength.  Each field depends on the superspace coordinates $(x^m,
\theta^\alpha_I)$. Vertex operators of this form were first discussed in
ten dimensions in \cite{Siegel}.

The vertex operator $\Phi_{\rm hybrid}$ needs to be compared with the
corresponding vertex operator in the pure spinor formalism. It is
important to note that, as for the hybrid string, we need to impose
both the pure spinor BRST condition (\ref{325}), $Q_{\rm ps} \Phi =
0$, as well as the physical state conditions (\ref{psc}), now
rewritten in terms of pure spinor variables. The condition (\ref{325})
alone does not put the theory on-shell. 

Consider for instance the massless compactification-independent open
string vertex operator which is obtained by solving $Q_{\rm ps} \Phi =
0$. In six dimensions it has the form
\begin{align}
\Phi_{\rm ps} = \lambda^\alpha_I A^I_\alpha(x,\theta^\alpha_I) \,,
\end{align}
where the ghost-number one spinor superfield $A^I_\alpha$ contains the
Yang-Mills degrees of freedom. As repeatedly stated \cite{Berkovits02,
Grassi:2005sb}, $Q_{\rm ps}$ only selects the {\em off-shell} field content
of six-dimensional $\N=1$ super-Yang-Mills.  $Q_{\rm ps} \Phi=0$
implies 
\begin{equation}
\lambda^\alpha_I \lambda^\beta_J D^I_\alpha A^J_\beta =0 \,, \label{cons}
\end{equation}
where $D_\alpha^I = \frac{\partial}{\partial \theta^\alpha_I} +
\frac{1}{2} \varepsilon^{IJ} (\gamma^m \theta_J)_\alpha \frac{\partial}
{\partial x^m}$.  

Since $\lambda_I^\alpha \lambda_J^\beta$ is a symmetric tensor under the
exchange of $(I,\alpha)$ and $(J,\beta)$, it projects $D^I_\alpha
A^J_\beta$ onto its symmetric part under this involution.  This part
decomposes as
\begin{align} \label{DA}
 D^I_{\alpha} A^J_{\beta}+ D^J_{\beta} A^I_{\alpha} =
 \varepsilon^{IJ} 
\Gamma^m_{\alpha\beta} A_m + ... \,,
\end{align} {\em i.e.}\ into a vector $A_m$ and other $n$-form
contributions indicated by ellipses. Note here that in six dimensions
a general antisymmetric bispinor $f_{\alpha\beta}$
($\alpha,\beta=1,...,4$) is related to a vector $A_m$ ($m=0,...,5$) by
$f_{\alpha\beta}=\Gamma^m_{\alpha\beta} A_m$. By substituting this
into (\ref{cons}) and using the pure spinor constraint
(\ref{constraints}), one can show that all $n$-form contributions
vanish, {\em i.e.}\ all terms in the ellipses in (\ref{DA}) are zero.
Then, (\ref{DA}) becomes exactly the linearised constraint
$F^{IJ}_{\alpha\beta}=0$, which is imposed on the superspace field
strength $F^{IJ}_{\alpha\beta}$, cf.\ with Eq.~(3.17) in \cite{Howe}.
Unlike in ten dimensions, this constraint is {\em off-shell} since one
cannot deduce the equation of motions from it. (Since $Q_{\rm
  hybrid}=Q_{\rm ps}$ this implies that also $Q_{\rm hybrid}$ selects
only the off-shell field content.) It is therefore natural to assume
that the conditions (\ref{psc}) put the pure spinor theory on-shell,
as it does in hybrid string theory.\footnote{A different method to put
  the pure spinor theory on-shell was proposed in \cite{Cederwall}.}

\medskip

In conclusion, we have shown that for superstring theory on
$\RR^6\times T^4$ a gauge-fixed version of the hybrid string is
related to a (hybrid) pure spinor string theory by a simple field
redefinition given by Eq.~(\ref{identification}). In particular, this
transformation identifies both the actions as well as the hybrid
string BRST-like condition (\ref{320}) and the pure spinor BRST
operator (\ref{325}). These BRST operators determine the off-shell
field content of $\N=1$ six-dimensional super Yang-Mills.  We argued
that in order to put the hybrid pure spinor theory on-shell a physical
vertex operator also has to satisfy the conditions (\ref{psc}). Of
course, for this the physical state conditions (\ref{psc}) must be
rewritten in pure spinor variables using the identifications
(\ref{identification}), which we have not done explicitly. We are
fairly optimistic though that the conditions (\ref{psc}) will then
provide the required equations of motion for the gauge field
$A^I_\alpha(x,\theta^\alpha_I)$, as it does for the corresponding
hybrid string vertex operator. The hybrid version of the pure spinor
string then provides the appropriate framework for six-dimensional
pure spinors.

\medskip

An open question is the relation of the {\em four}-dimensional pure
spinor action to the corresponding hybrid string on $\RR^4 \times T^6$
\cite{Berkovits:1994wr}. Here we encounter a puzzle
\cite{Berkovits07}: The central charge of the pure spinor theory
parameterizing the part on $\RR^4$ is zero \cite{Grassi:2005sb,
  Wyllard}.  If~we wish to describe the six internal directions in RNS
variables, we obtain a (topological) $\N=2$ string with $\hat c=3$
($c=9$).  However, a critical $\N=2$ string has $\hat c=2$ ($c=6$),
and the pure spinor theory cannot be related to hybrid strings in a
simple way. Possibly such a naive compactification of the pure spinor
theory describes the BPS sector of the compactified superstring
\cite{B3}. More work is needed here to make the relation precise.

\section*{Acknowledgements} 

We would like to thank Matthias Gaberdiel, Stefan Hohenegger, Peter
R\o nne, Volker Schomerus and Niclas Wyllard for useful discussions
related to this work. We are also grateful to Nathan Berkovits 
for helpful comments on a preliminary version of this paper.
This research has been supported by the Swiss
National Science Foundation.

\appendix

\section*{Appendix}

\setcounter{equation}{0}
\section{Six-dimensional gamma matrices}\label{appA}

We give an explicit realisation of the matrices $\Gamma^m_{\alpha\beta}$ used
throughout this paper satisfying (\ref{Clifford}). They can be chosen to be
\begin{equation*}
\begin{array}{lcl}
 \Gamma^0 = -i\mathds{1} \otimes \tau^2 && \Gamma^3=\tau^2 \otimes \tau^3 \\
 \Gamma^1 = \tau^2 \otimes \tau^1 && \Gamma^4=-i \tau^2 \otimes \mathds{1} \\
 \Gamma^2 = i\tau^1 \otimes \tau^2 && \Gamma^5=-i\tau^3 \otimes \tau^2 \,,\\
\end{array}
\end{equation*}
where $\tau^i$ are the usual two-dimensional Pauli matrices. The Weyl indices
are raised by the epsilon tensor according to the rule
\begin{equation*}
 \left(\Gamma^m\right)^{\alpha\beta}=\frac{1}{2}\varepsilon^{
\alpha\beta\gamma\delta } \Gamma^m_{\gamma\delta} \,.
\end{equation*}
A useful identity is
\begin{equation}
(\Gamma^m)_{\alpha\beta}(\Gamma_m)_{\gamma\delta}=-2\varepsilon_{
\alpha\beta\gamma\delta} \,. \label{identity}
\end{equation}

\setcounter{equation}{0}
\section{Summary of fields in RNS, hybrid and pure spinor 
formalism}\label{appB}

In this appendix we summarize the fields occurring in the three
worldsheet formalisms of string theory on $\RR^6 \times T^4$.

In the {\em RNS formalism} the external part on $\RR^6$ is described
in terms of six bosons and fermions, $x^m$ and $\psi^m$ ($m=0,...,5$),
while the internal part on $T^4$ is parameterized by four bosons and
fermions, $Y^i$ and $\eta^i$ ($i=1,...,4$). The ghost sector is given
by the standard $bc$ and $\beta\gamma$ systems. These fields and their
properties are listed in the upper left box of table~\ref{tab1}. 

\begin{table}[ht]
\begin{center}
\begin{tabular}{|c|c|c|c|c|}
 \hline RNS&&$c$ &$h$ & $\varepsilon$\\ \hline 6 bosons & $x^m$ & $6$
 & $1$ & $-1$\\ 6 fermions & $\psi^m$ & $3$ & $\frac{1}{2}$ & $+1$\\
 \hline 2 bosons & $\beta,\gamma$ & $11$ & $\frac{3}{2},-\frac{1}{2}$
 & $-1$\\ 2 fermions & $b,c$ & $-26$ & $2,-1$ & $+1$\\ \hline $T^4$ &
 $Y^i$, $\eta^i$ & $6$ & $1,\frac{1}{2}$ & $\mp 1$\\ \hline
\end{tabular} $\quad\longrightarrow\quad$
\begin{tabular}{|c|c|c|c|c|}
\hline
 hybrid&& $c$&$h$& $\varepsilon$\\
 \hline
 6 bosons & $x^m$ & $6$ & $1$ & $-1$\\
 8 fermions & $p_\alpha, \theta^\alpha$ & $-8$ & $1, 0$ & $+1$\\
 \hline
 1 boson & $\rho$ & $28$ & $0$ & $-1$\\
 1 boson & $\sigma$ & $-26$ & $0$ & $+1$\\
 \hline
 $T^4$ & $Y^i$, $\eta^i$ & $6$ & $1,\frac{1}{2}$& $\mp 1$\\
 \hline
\end{tabular} \\
\medskip
\hspace{3.5cm} $\downarrow$\\
\medskip
\begin{tabular}{|c|c|c|c|c|}
\hline
 pure spinor && $c$&$h$ & $\varepsilon$\\
 \hline
 6 bosons & $x^m$ & $6$ & $1$ & $-1$\\
 16 fermions & $p^I_\alpha, \theta^\alpha_I$ & $-16$ & $1, 0$ & $+1$\\
 \hline
 5 pure spinors & $\lambda^\alpha_I, w^I_\alpha$ & $10$ & $0,1$ & $-1$ \\
 \hline
 $T^4$ & $Y^i$, $\eta^i$ & $6$ & $1,\frac{1}{2}$& $\mp 1$\\
 \hline
\end{tabular}
\end{center}
\caption{Overview of the fields in the RNS, hybrid and pure spinor formalism 
of string theory on $\RR^6 \times T^4$. $c$ and $h$ denote the
contribution to the total central charge and the conformal weight of
the fields.  The value $\varepsilon=+1$ ($\varepsilon=-1$) refers to
Fermi (Bose) statistics.} \label{tab1}
\end{table}

The {\em hybrid formalism} is obtained from the RNS formalism
\cite{BVW} by first embedding the critical $\N=1$ RNS string 
into a critical $\N=2$ string and then, after twisting, performing the
following field redefinition. The bosons $x^m$ are the same as in the
RNS string. The fermions and ghosts are reorganised into eight
fermions $p_\alpha,
\theta^\alpha$ $(\alpha=1,...,4)$ and two chiral bosons, $\sigma$ 
and~$\rho$. The latter are obtained by bosonising both the $bc$ as
well as the $\beta\gamma$ system in the standard way, i.e.\ as
$(b,c)=(e^{- i \sigma}, e^{i \sigma})$ and $(\beta,\gamma)=(e^{-
\phi+\kappa} \partial\kappa, e^{\phi-\kappa})$.  The corresponding
background charges are $Q_\kappa=1$, $Q_\phi=2$ and $Q_\sigma=3$
($\varepsilon_\kappa=-\varepsilon_\phi=1$).
Then $\sigma$ and $\rho$ are defined by
\begin{align}
\partial \sigma = i bc \,,\qquad \rho=-2\phi-i\kappa-i H^{RNS}_C \,,
\end{align}
where $H^{RNS}_C=H_4+H_5$ are the bosonised fermions of $T^4$. Both
fields are spacetime bosons of conformal weight zero but have opposite
statistics, $\varepsilon_\sigma=-\varepsilon_\rho=1$. Their
contribution to the central charge is $c=1-\varepsilon 3Q^2$ with
background charges $Q_\sigma=Q_\rho=3$, and therefore $c_\sigma=-26$
and $c_\rho=28$.  The fermions $p_\alpha,
\theta^\alpha$ are defined in terms of the RNS variables as
\begin{align}
\theta^\alpha &= [e^{\frac{1}{2}\phi} \Sigma^{+++++},
e^{\frac{1}{2}\phi} \Sigma^{--+++}, e^{\frac{1}{2}\phi} \Sigma^{-+-++}, 
e^{\frac{1}{2}\phi} \Sigma^{+--++}] \,,\nn\\
p_\alpha &= [e^{-\frac{1}{2}\phi} \Sigma^{-----},
e^{-\frac{1}{2}\phi} \Sigma^{++---}, e^{-\frac{1}{2}\phi} \Sigma^{+-+--}, 
e^{-\frac{1}{2}\phi} \Sigma^{-++--}]\,,
\end{align}
where $\phi$ is the $\beta\gamma$-boson and $\Sigma^\alpha$~is the
spin field of conformal weight $\frac{5}{8}$ defined by
\begin{align}
\Sigma^\alpha= e^{\frac{i}{2} \sum^5_{I=1} \epsilon_I H_I} \,,
\end{align}
with $\epsilon_I=\pm 1$. The bosons $H_{1,2,3}$ and $H_{4,5}$ are
obtained by bosonising the fermions $\psi^m$ and $\eta^i$,
respectively. Since $e^{n\phi}$ has weight $-\frac{n^2}{2}-n$, which
is $-\frac{5}{8}$ and $\frac{3}{8}$ for $e^{\pm \phi/2}$, $p_\alpha$
and $\theta^\alpha$ form four (fermionic) $bc$-systems with weights
$(1,0)$. Their contribution to the central charge is
$c=4\cdot(-2)=-8$.\footnote{For $bc$ ($\beta\gamma$) systems the
anomaly contribution of the ghosts is
$c=-2\varepsilon(6\lambda(\lambda-1)+1)$, where $\lambda=h_b$
($h_\beta$) and $\varepsilon=1$ ($-1$).} The internal part on $T^4$ is
the same as in the RNS formalism. The fields of the hybrid formalism
are summarized in the upper right box of
table~\ref{tab1}.\footnote{The table lists only the matter part of the
$\N=2$ hybrid string. Note that the critical central charge for the
matter part of an $\N=2$ string is $c=6$.}

The {\em pure spinor formalism} requires again six bosons $x^m$ and
now sixteen fermions $p^I_\alpha$ and $\theta^\alpha_I$ ($I=1,2;
\alpha=1,...,4$), twice as many as in the hybrid formalism.
The pure spinor part consists of the fields $\lambda^\alpha_I,
w^I_\alpha$, which because of the pure spinor condition formally
represent five (bosonic) $\beta\gamma$-systems with weights $(0,1)$.
Their contribution to the central charge is therefore $c=5\cdot2=10$.
The internal part is again as in the RNS formalism. The connection
between the hybrid and the pure spinor formalism is described in
section~\ref{sec3}. The fields and their properties are shown in the
lower box of table~\ref{tab1}.



\begin{thebibliography}{99}

\bibitem{B1}
  N.~Berkovits, {\it
  Super-Poincare covariant quantization of the superstring},
  JHEP {\bf 0004}, 018 (2000)
  [arXiv:hep-th/0001035].

\bibitem{B2}
  N.~Berkovits,
  {\it Multiloop amplitudes and vanishing theorems using the pure spinor
  formalism for the superstring},
  JHEP {\bf 0409}, 047 (2004)
  [arXiv:hep-th/0406055].

\bibitem{B3}
  N.~Berkovits, {\it
  Pure spinor formalism as an N = 2 topological string},
  JHEP {\bf 0510} (2005) 089
  [arXiv:hep-th/0509120].

\bibitem{Berkovits02} 
  N.~Berkovits,
  {\it ICTP lectures on covariant quantization of the superstring},
  arXiv:hep-th/0209059.

\bibitem{Oz:2009tb}
  Y.~Oz,
  {\it The Pure Spinor Formulation of Superstrings},
  Class.\ Quant.\ Grav.\  {\bf 25}, 214001 (2008)
  [arXiv:0910.1195 [hep-th]].

\bibitem{Bedoya:2009np}
  O.~A.~Bedoya and N.~Berkovits,
  {\it GGI Lectures on the Pure Spinor Formalism of the Superstring},
  arXiv:0910.2254 [hep-th].

\bibitem{Grassi:2005sb}
  P.~A.~Grassi and N.~Wyllard,
  {\it Lower-dimensional pure-spinor superstrings},
  JHEP {\bf 0512} (2005) 007
  [arXiv:hep-th/0509140].

\bibitem{Wyllard}
  N.~Wyllard, {\it
  Pure-spinor superstrings in d = 2, 4, 6},
  JHEP {\bf 0511} (2005) 009
  [arXiv:hep-th/0509165].

\bibitem{Adam:2006bt}
  I.~Adam, P.~A.~Grassi, L.~Mazzucato, Y.~Oz and S.~Yankielowicz,
  {\it Non-critical pure spinor superstrings},
  JHEP {\bf 0703}, 091 (2007)
  [arXiv:hep-th/0605118].

\bibitem{Berkovits07}
  N.~Berkovits,
  {\it Explaining the Pure Spinor Formalism for the Superstring},
  JHEP {\bf 0801}, 065 (2008)
  [arXiv:0712.0324 [hep-th]].

\bibitem{BVW}
  N.~Berkovits, C.~Vafa and E.~Witten, {\it
  Conformal field theory of AdS background with Ramond-Ramond flux},
  JHEP {\bf 9903}, 018 (1999)
  [arXiv:hep-th/9902098].

\bibitem{Berkovits:1999du}
  N.~Berkovits, {\it
  Quantization of the type II superstring in a curved six-dimensional
  background},
  Nucl.\ Phys.\  B {\bf 565} (2000) 333
  [arXiv:hep-th/9908041].

\bibitem{Berkovits:1999xv}
  N.~Berkovits, {\it
  Quantization of the superstring in Ramond-Ramond backgrounds},
  Class.\ Quant.\ Grav.\  {\bf 17}, 971 (2000)
  [arXiv:hep-th/9910251].

\bibitem{Henneaux:1992ig}
  M.~Henneaux and C.~Teitelboim, {\it
  Quantization of gauge systems},
{\it  Princeton, USA: Univ. Pr. (1992) 520 p}.

\bibitem{Theisen}
  J.~Kappeli, S.~Theisen and P.~Vanhove,
  {\it Hybrid formalism and topological amplitudes},
  arXiv:hep-th/0607021.

\bibitem{Siegel}
  W.~Siegel,
  {\em Classical Superstring Mechanics,}
  Nucl.\ Phys.\  B {\bf 263}, 93 (1986).

\bibitem{Howe}
  P.~S.~Howe, G.~Sierra and P.~K.~Townsend,
  {\it Supersymmetry In Six-Dimensions},
  Nucl.\ Phys.\  B {\bf 221}, 331 (1983).

\bibitem{Cederwall}
  M.~Cederwall and B.~E.~W.~Nilsson,
  {\it Pure Spinors and D=6 Super-Yang-Mills},
  arXiv:0801.1428 [hep-th].

\bibitem{Berkovits:1994wr}
  N.~Berkovits,
  {\it Covariant quantization of the Green-Schwarz superstring in a Calabi-Yau
   background},
  Nucl.\ Phys.\  B {\bf 431}, 258 (1994)
  [arXiv:hep-th/9404162].


\end{thebibliography}
\end{document}